\documentstyle[12pt]{article} 

\newcommand{\EQ}{\begin{equation}}
\newcommand{\EN}{\end{equation}}
\newcommand{\bea}{\begin{eqnarray}} 
\newcommand{\ena}{\end{eqnarray}}
\newcommand{\vs}[1]{\vspace{#1 mm}}

\renewcommand{\a}{\alpha}
\renewcommand{\b}{\beta}
\renewcommand{\c}{\gamma}
\renewcommand{\d}{\delta}
\newcommand{\e}{\epsilon}
\newcommand{\C}{\Gamma}
\def\bbox{{\,\lower0.9pt\vbox{\hrule \hbox{\vrule height 0.2 cm
\hskip 0.2 cm \vrule height 0.2 cm}\hrule}\,}}
\newcommand{\dsl}{\pa \kern-0.5em /}

\newcommand{\pa}{\partial}
\renewcommand{\t}{\theta}
\renewcommand{\k}{\kappa}

\def\apr{{a^\prime}}
\def\bpr{{b^\prime}}
\def\cpr{{c^\prime}}
\def\dpr{{d^\prime}}
\def\alpr{{\alpha^\prime}}
\def\bepr{{\beta^\prime}}
\def\gapr{{\gamma^\prime}}

\def\aha{{\hat{a}}}

\newcommand{\nn}{\nonumber\\}
\newcommand{\p}[1]{(\ref{#1})}

\begin{document}

\topmargin 0pt
\oddsidemargin 0mm
\renewcommand{\thefootnote}{\fnsymbol{footnote}}
\begin{titlepage}

\setcounter{page}{0}
\begin{flushright}
hep-th/9906013
\end{flushright}

\vs{10}
\begin{center}

{\Large\bf Super 0-brane and GS Superstring Actions on $AdS_2 \times S^2$}

\vs{15}

{\large 
Jian-Ge Zhou\footnote{e-mail address: jgzhou@het.phys.sci.osaka-u.ac.jp}}

\vs{10}
{\em Department of Physics, Osaka University,
Toyonaka, Osaka 560-0043, Japan} \\
\end{center}

\vs{15}
\centerline{{\bf{Abstract}}}
\vs{5}

The super 0-brane and GS superstring actions on AdS$_2 \times S^2$ 
background with 2-form flux are constructed by supercoset approach.
We find the super 0-brane action contains two parameters which
are interpreted as the electric and magnetic charges of the
super 0-brane. The obtained super 0-brane action describes
the BPS saturated dyonic superparticle moving on AdS$_2 \times S^2$
background. The WZ action contains the required coupling with 
2-form flux. For GS superstring, we find the string action on 
AdS$_2 \times S^2$ takes the same form as those in AdS$_3 \times S^3$
and AdS$_5 \times S^5$ with RR field background.

\end{titlepage}
\newpage

\renewcommand{\thefootnote}{\arabic{footnote}}
\setcounter{footnote}{0}

{\Large\bf 1.Introduction}
\vs{5}

Recently, motivated by the AdS/CFT correspondence
~\cite{M1}-\cite{W1}, there has arisen renewed interest
to study D-brane and string dynamics in curved space. The
type IIB Green-Schwarz (GS) superstring action was constructed in 
AdS$_5 \times S^5$ background in terms of supercoset formalism
~\cite{MT1}. This action possesses global $SU(2,2\mid 4)$
super-invariance,  has 
$\k$-symmetry and 2D reparametrization
invariance as its local symmetries, and reduces to the conventional
type IIB GS superstring action in the flat background
limit. The other related construction for GS superstring, super 
D3-brane, D1-brane
on AdS$_5 \times S^5$, and super M-branes on  AdS$_4 \times S^7$
and AdS$_7 \times S^4$ have been discussed in~\cite{KRR}-\cite{O1}.
The gauge-fixing of $\k$-symmetry was carried out in two
different approaches: the supersolvable algebra approach~\cite{DF} and
Killing gauge approach~\cite{K1}, and it was shown that the
gauge-fixed actions obtained in the two approaches agree through
appropriate rearrangement of fields~\cite{P1}.

Even the above studies have made much progress on the
understanding to D-brane and string dynamics in AdS$_{p+2} \times S^{D-p-2}$
background (where p is the dimensions of p-branes or strings, and D is 
the spacetime dimensions), the $\k$-symmetry gauge-fixing and
quantization seem
still to pose some difficulties~\cite{PST, RR, P2}, then it would be 
appropriate
to study the D-branes and superstring propagation on some simple
Ramond-Ramond (RR) backgrounds. One of the simple backgrounds is type IIB 
string on AdS$_3 \times S^3 \times M_4$ ($M_4$ can be chosen as K3 or
$T^4$), which is the near-horizon geometry of type IIB D1-D5 brane
configuration~\cite{M1}. The GS superstring and D-brane actions on 
AdS$_3 \times S^3$ has been constructed in~\cite{P3} and~\cite{S1}. 
However, there
is a much more simpler background AdS$_2 \times S^2 \times M_6$ 
(where $M_6$ is a Calabi-Yau manifold or $K3 \times T^2$ or  $T^6$),
which can be obtained, for example, from the near-horizon geometry of 
$3 \perp 3 \perp 3
\perp 3$ configuration in type IIB string theory~\cite{M1} . In
D=4, N=2 supergravity, the AdS$_2 \times S^2$ background
is the Bertotti-Robinson (BR) solution ~\cite{br1} which is the 
near-horizon geometry of 3+1  dimensional Reissner-Nordstr\"om black
hole. Therefore, it is quite interesting to see how p-branes and
GS superstring propagate on AdS$_2 \times S^2$ background
with 2-form flux.

On the other hand, it was shown in  ~\cite{CDK} that radial motion of
a superparticle with zero angular momentum near the horizon
of an extreme Reissner-Nordstr\"om black hole (AdS$_2 \times S^2$)
is described by an $Osp(1 \mid 2)$-invariant superconformal mechanics,
and it was further argued in ~\cite{CDK} that the full superparticle
dynamics should be invariant under the larger $SU(1,1 \mid 2)$
superconformal group because this is the superisometry group of 
AdS$_2 \times S^2$. This full dynamics describe not only the radial
motion of the superparticle but also its motion on $S^2$. In ~\cite{AIB},
the full  $SU(1,1 \mid 2)$-invariant action was constructed from an
extension to N=4 superconformal mechanics in the worldline superfield
formalism (analogous to the worldline supersymmetry version). Then
one may ask whether it is possible to construct the full 
$SU(1,1 \mid 2)$-invariant action  in spacetime supersymmetry version,
which possesses local $\k$-symmetry and one-dimensional 
reparametrization invariance.

Due to the above motivations, in this paper we construct super
0-brane and GS superstring actions on AdS$_2 \times S^2$ background
by supercoset approach. The action is built out of the Cartan 1-forms
$L^a$, $L^{\apr}$ and $L^I$. 
For super 0-brane, we find the action with 
two free parameters A and B (their definition will be given below)
where the parameter A can be explained as the electric charge of the
super 0-brane, and B corresponds to its magnetic charge.
And by comparing with the results in ~\cite{CDK, AIB},
we find that $L^a$ describes the radial motion part of the super
0-brane, and $L^{\apr}$  corresponds to its motion on $S^2$.
The obtained super 0-brane
action has global $SU(1,1 \mid 2)$ super-invariance, and is
invariant under local  $\k$-symmetry and one-dimensional 
reparametrization invariance. The expression for 
$\k$-symmetry includes two parameters A and B.  In the present
construction, 
the mass of super 0-brane is required to be $m = {(A^{2} + B^{2})}^{1/2}$
by  $\k$-symmetry, which means the action, in some sense, describes
the BPS saturated dyonic superparticle moving on  AdS$_2 \times S^2$ 
background. Especially, the obtained
WZ action  contains the $r^{-2}$ term which is expected from the
conformal mechanics model ~\cite{AFF}. For GS superstring, we
find that the string  action  on AdS$_2 \times S^2$ with 2-form flux
takes the same form as those in AdS$_3 \times S^3$ and
AdS$_5 \times S^5$ with RR field background.

The layout of the paper is as follows. In section 2, 
the structure of superalgebra  $SU(1,1 \mid 2)$ is presented and 
the corresponding Maurer-Cartan equation for the Cartan 1-forms is given.
In section 3, the super 0-brane action on AdS$_2 \times S^2$ is
constructed, its $\k$-symmetry is verified. Some properties
of the obtained 0-brane action are discussed. In section 4, GS
superstring action is constructed, and its $\k$-symmetry is
analysed. In section 5, we present our summary and discussions.

\vs{5}
{\Large\bf 2. $SU(1,1 \mid 2)$ superalgebra and Maurer-Cartan equation
for Cartan 1-forms}
\vs{5}

The algebra of isometry of AdS$_2 \times S^2$ background is given
by  $SU(1,1 \mid 2)$ , its bosonic subalgebra consists of 
$SO(1,2) \oplus SO(3)$. The 4-dimensional gamma matrices
and charge conjugation matrix can be decomposed into \footnote{We use
the notation and convention in~\cite{MT1}.}
\bea
\C^a= \c^a \otimes 1,\;\;
\C^{\apr} = \c  \otimes \c^{\apr}, \;\;
{\cal C}= C \otimes C^\prime
\label{dec}
\ena
where $a = 0, 1$, ${\apr} = 2, 3$, $\c=\c_0 \c_1$, and we have chosen the 
four-dimensional spinors as Majorana ones. Observe that C is
antisymmetric, $C^\prime$ is symmetric, so that $\cal C$ is 
antisymmetric. The symmetric matrices are $C\c^a, C\c^{ab}, C^\prime,
C^\prime\c^{\apr}$, and the antisymmetric ones are C,
$C^\prime\c^{\apr\bpr}$~\cite{KT}. 
And we define $\C_5=\C_0\C_1\C_2\C_3=\c\otimes\gapr$,
with $\gapr=\c^2\c^3$. The charge conjugation $\psi_c$ of a spinor  
 $\psi$ is $\psi^{\rm T} {\cal C} \equiv {\bar\psi} =
{\psi^{\dag}}\C^0$. The fermionic generators $Q_{\a\alpr I}$ are 
Majorana spinors with $\a=1,2$, $\alpr=1,2$,  $I=1,2$. Then $SU(1,1 \mid 2)$
supersymmetry algebra is given by
\bea
\left[P_a,P_b \right]&=&J_{ab}\nn 
\left[P_\apr,P_\bpr \right]&=&-J_{\apr\bpr}\nn 
\left[P_a,J_{bc} \right]&=&\eta_{ab}P_c-\eta_{ac}P_b\nn
\left[P_\apr,J_{\bpr\cpr} \right]&=&\eta_{\apr\bpr}P_\cpr
-\eta_{\apr\cpr}P_\bpr\\
\left[J_{ab},J_{cd} \right]&=&0 \nn 
\left[J_{\apr\bpr},J_{\cpr\dpr} \right]&=&0 
\label{bb2}
\ena
\bea
\left[Q_{\a \alpr I}, P_a \right]&=&\frac{{\rm 1}}{2}\e_{{IJ}}
Q_{\b \alpr J}(\c_{a} \c)_{\b \a} \nn 
\left[Q_{\a \alpr I}, P_\apr \right]
&=&-\frac{1}{2}\e_{{IJ}} Q_{\a \bepr J} (\c_\apr)_{\bepr \alpr}\nn
\left[Q_{\a \alpr I},J_{ab} \right]&=&-\frac{1}{2} Q_{\b\alpr
I}(\c_{ab})_{\b\a} \nn
\left[Q_{\a \alpr I},J_{\apr\bpr}\right ]&=&-\frac{1}{2}Q_{\a\bepr
I}(\c_{\apr\bpr})_{\bepr\alpr}
\label{bf}
\ena
\bea
\{Q_{\a \alpr I }, Q_{\b \bepr J } \}
&=&\d_{{IJ}}
\bigg[2(C\c_a)_{\a\b}C_{\alpr\bepr} P_a
+2(C\c)_{\a\b}(C^\prime\c^\apr)_{\alpr\bepr}
P_\apr\bigg] 
\nn 
&+&\e_{{IJ}}
\bigg[-(C\c\c^{ab})_{\a\b} C_{\alpr\bepr}J_{ab}
+(C\c)_{\a\b}(C^\prime\c^{\apr\bpr})_{\alpr\bepr}
J_{\apr\bpr}\bigg] \ .
\label{ff}
\ena

Since $a=0,1, \apr=2,3$, there are only $J_{01}, J_{23}$ nonzero, we
put the right side of Eq.\p{bb2} to be zero. Note that both sides of 
Eq.\p{ff} are symmetric under the exchange of of $(\a\alpr I)
\leftrightarrow (\b\bepr J)$. Here we would like to mention that
if we choose the decomposion given by
\bea
\C^a= \c^a \otimes \gapr,\;\;
\C^{\apr} = 1 \otimes \c^{\apr}, \;\;
{\cal C}= C \otimes C^\prime
\ena
instead of \p{dec}, the $SU(1,1\mid 2)$ supersymmetry algebra takes
slightly different form to the above, but they are equivalent.

In the scaling limit $P_{\aha}\rightarrow
RP_{\aha}$,  $J\rightarrow J$ and 
$Q^I\rightarrow \sqrt{R}Q^J$ with $R \rightarrow \infty$ (R is the 
radius of AdS$_2 \times S^2$), the $SU(1,1\mid 2)$
superalgebra is reduced to the supersymmetric algebra in flat 4D
spacetime.

The left-invariant Cartan 1-forms 
\bea
L^A=dX^{M}L^{A}_M, \;\; 
X^{M} = (x, \theta)
\ena
satisfy the following Mauer-Cartan equation
\bea
dL^a&=&- L^b\wedge L^{ba}
- L^{\a\alpr I}(C\c^a)_{\a\b}C_{\alpr\bepr}\wedge L^{\b\bepr I}\nn
dL^{a'}&=&-L^{\bpr}\wedge L^{b'a'}
- L^{\a\alpr I}(C\c)_{\a\b}(C^\prime\c^{\apr})_{\alpr\bepr}\wedge 
L^{\b\bepr I}\nn
dL^{ab}&=&-L^a\wedge L^b+\e_{IJ}L^{\a\alpr I}(C\c\c^{ab})_{\a\b}
C_{\alpr\bepr}\wedge L^{\b\bepr J}\nn
dL^{a'b'}&=&L^{a'}\wedge L^{b'}
-\e_{IJ} L^{\a\alpr I}(C\c)_{\a\b}(C^\prime\c^{\apr\bpr})_{\alpr\bepr}\wedge 
L^{\b\bepr J}\nn
dL^I&=& \frac{1}{2}\e^{IJ} \c^{a}\c {L^J}\wedge L^a
-\frac{1}{2}\e^{IJ} \c^{\apr}{L^J}\wedge L^{\apr}
\nn 
&+& \frac{1}{4}\c^{ab}{L^I}\wedge L^{ab}
+\frac{1}{4}\c^{\apr\bpr}L^I\wedge L^{\apr\bpr}
\label{mc}
\ena

From the rescaling of the generators and Eq.\p{mc}, the Cartan
1-forms transforms as 
\bea
L^a\rightarrow R^{-1}L^a,  \;\; 
L^{\apr}\rightarrow R^{-1}L^{\apr} \;\;
L^{ab}\rightarrow L^{ab},\;\;
L^{\apr\bpr}\rightarrow L^{\apr\bpr},\;\;
L^I\rightarrow R^{-1/2}L^I
\label{ls}
\ena

\vs{5}
{\Large\bf 3. Super 0-brane action on $AdS_2 \times S^2$}
\vs{5}

The general structure of the super 0-brane action on 
Ad$S_2 \times S^2$ background in terms of supercoset formalism can 
be written as

\bea
I_{0-brane} = \int\limits_{\pa {\cal M}_2} {{\cal L}_{DBI}} +
\int\limits_{{\cal M}_2}{{\cal L}_{WZ}}\nn
{\cal L}_{DBI} = -m\sqrt {-({L^a_0}{L^a_0} +
{L^{\apr}_0}{L^{\apr}_0})}
\ena
where m is the mass of super 0-brane, and
\bea
L^{\aha}_0 = ({\frac{dX^M}{dt}})L^{\aha}_M,\;\; 
L^I_0 = ({\frac{dX^M}{dt}})L^I_M
\ena
Since ${\cal L}_{WZ}$ is a closed 2-form, we require that
\bea
d{\cal L}_{WZ} = 0
\ena

Since under the action of an arbitrary element of the isometry
group the vielbeins transform as tangent vectors of the stability
subgroup, ${\cal L}_{WZ}$ should be invariant of the subgroup
$SO(1,1)\otimes SO(2)$ in  Ad$S^2 \times S^2$ case. The only
form for ${\cal L}_{WZ}$ built out of $L^a, L^{\apr}$ and $L^I$,
which can be reduced to flat case in the limit $R\rightarrow \infty$,
is given by\footnote{We can also add other terms to  ${\cal L}_{WZ}$, but
they cannot be reduced properly in flat limit, which we shall discuss
below.}

\EQ
{\cal L}_{WZ} = {\cal H}_{1} + {\cal H}_{2}
\EN
with
\bea
{\cal H}_{1} &= &A\e^{IJ}L^{\a\alpr I}C_{\a\b}C_{\alpr\bepr}\wedge 
L^{\b\bepr J}
+ {A^\prime} \e_{ab}L^a\wedge L^b\nn
{\cal H}_{2} &= &B\e^{IJ}L^{\a\alpr I}(C\c)_{\a\b}
(C^\prime\c^\prime)_{\alpr\bepr}\wedge 
L^{\b\bepr J}
+ B^\prime \e_{\apr\bpr}L^{\apr}\wedge L^{\bpr}
\ena
where we have defined $\e^{01} = -\e_{01} = 1, \e^{23} = \e_{23} = 1$.
From the Maurer-Cartan equations, one finds
\bea
d{\cal H}_{1}& =& A\bar L^{I}\c_{a}\c\wedge L^{I}\wedge L^a
+ 2A^\prime \bar L^{I}\c_{a}\c\wedge L^{I}\wedge L^a\nn
d{\cal H}_{2}& =& B\bar L^{I}(\c\otimes\c_{\apr}\gapr)\wedge L^{I}
\wedge L^{\apr}
+ 2B^\prime \bar L^{I}(\c\otimes\c_{\apr}\gapr)\wedge L^{I}\wedge L^{\apr}
\ena
where we have used $\c_a\c = \e_{ab}\c^b, \c_{\apr}\gapr =
\e_{\apr\bpr}\c^{\bpr}$.

To demand $d{\cal L}_{WZ} = 0$, the only choice is
\bea
A^\prime = - A/2, \;\; 
B^\prime = - B/2
\ena
but no further restriction between A and B, and we have
\bea
d{\cal H}_{1} = 0, \;\; d{\cal H}_{2}= 0
\ena
Then the super 0-brane action on Ad$S_2 \times S^2$ can be
written as
\bea
I_{0-brane}&= &-m\int{dt \sqrt {-({L^a_0}{L^a_0} +
{L^{\apr}_0}{L^{\apr}_0})}}\nn
&& 
+ \int\limits_{{\cal M}_2}\{A\e_{IJ}{\bar L}^I \wedge L^{J}
+ B\e_{IJ}{\bar L}^{I}\C_{5} \wedge L^{J}\nn
&& 
- \frac{A}{2}\e_{ab}L^a \wedge L^b - 
\frac{B}{2}\e_{\apr\bpr}L^{\apr}\wedge L^{\bpr}\}
\label{ab}
\ena

In \p{ab}, there are three parameters m, A and B, where m is the mass
of super 0-brane, the relation among m, A and B will be determined
by $\k$-symmetry.

\vs{2}

When we define $\delta x^a\equiv \delta X^M L^a_M$, 
$\delta x^{a'}\equiv \delta X^M L^{a'}_M$, and $ \delta\theta^I\equiv
\delta X^M L^I_M$, from \p{mc} we have in the variation of $\d\theta^I$
\bea
\d L^a&=&2\bar{L}^I \c^a \d\theta^I\nn
\d L^{\apr}&=&2\bar{L}^I \c\otimes\c^{\apr} \d\theta^I\nn
\d L^I&=&d\d \theta^I+\frac{1}{2}\e^{IJ}(L^a\c^a\c - 
L^{\apr}\c^{\apr})\d\theta^J\nn
&&
+ \frac{1}{4}(L_{ab}\c^{ab} 
+ L_{\apr\bpr}\c^{\apr\bpr})\d\theta^I
\label{vari}
\ena
The variation of ${\cal L}_{DBI}$ and ${\cal L}_{WZ}$  can be
got from \p{vari}

\bea
\d {\cal L}_{DBI}& =& \frac{2m \bar{L}^{I}_{0}L^{\aha}_{0}\C^{\aha}\d\t^I}
{\sqrt{-{L^{\aha}_{0}}L^{\aha}_{0}}}\nn
\d {\cal L}_{WZ}& =&-2d\biggr[A\e^{IJ}\bar{L}^{I}\d\t^J 
+ B\e^{IJ}\bar{L}^{I}\C_{5}\d\t^J \biggr]
\label{va}
\ena
Then the $\k$-symmetry transformation can be defined by 
\bea
\d_{\k}x^a = 0, \;\;
\d_{\k}x^{\apr} = 0,\nn
\d_{\k}\t^I = [(1+\C)\k)]^I
\label{dx}
\ena
with
\bea
\C  = {\frac{(A - B\c\otimes\gapr)(L^{a}_{0}\c^{a} +
L^{\apr}_{0}\c\otimes\c^{\apr})}{\sqrt{-(A^{2}+B^{2})
(L^{a}_{0}L^{a}_{0} + L^{\apr}_{0}L^{\apr}_{0})}}}{\cal E}
\label{pro}
\ena
\bea
{\cal E}= \pmatrix{
0  & -1 \cr 1 & 0 \cr }
\ena
\EQ
m = \sqrt {A^{2}+B^{2}}
\label{ma}
\EN
where the expression for $\k$-symmetry includes the parameters A and B,
and $m = \sqrt {A^{2}+B^{2}}$ occurs as a consequence of $\k$-symmetry
of super 0-brane action.

The projection $\C$ satisfies
\bea
\C^2 = 1, \;\; tr\C = 0
\ena
Then the super 0-brane action on Ad$S_2\times S^2$ can be recast into
\bea
I_{0-brane} = \int dt{\cal L}_{DBI} + \int\limits_{{\cal M}_2}{({\cal
L}_{WZ1} + {\cal L}_{WZ2})}
\label{ab1}
\ena
with
\bea
{\cal L}_{DBI} &=& -{(A^{2} + B^{2})}^{1/2}\sqrt{-(L^{a}_{0}L^{a}_{0}
+ L^{\apr}_{0}L^{\apr}_{0})}\nn 
{\cal L}_{WZ1}& = &A\e^{IJ}\bar{L}^I\wedge L^{J} + B\e^{IJ}\bar{L}^I
\C_{5}\wedge L^{J}\nn
{\cal L}_{WZ2}& =& -\frac{A}{2}\e_{ab}L^{a}\wedge L^{b}-
\frac{B}{2}\e_{\apr\bpr}L^{\apr}\wedge L^{\bpr}
\label{wz}
\ena
One can easily check that under the $\k$-symmetry transformation
\p{dx} with \p{pro} - \p{ma}, $I_{0-brane}$ is invariant, i.e.,
\bea
\d_{k}I_{DBI} + \d_{k}I_{WZ} = 0
\ena

Now we have got the super 0-brane action on Ad$S_2 \times S^2$
which is given by \p{ab1} and \p{wz}. This action has global $SU(1,1\mid 2)$
invariance, and has $k$-symmetry as well as one-dimensional
reparametrization invariance as its local symmetry. 

To see whether the above action can be reduced to the conformal
mechanics
model in ~\cite{AFF} when we switch off the fermionic part, let us 
consider the particle of mass m moving radially on the background
of BR solution in D=4, N=2 supergravity, which is given by~\cite{GH, FKS}
\bea
ds^2&=&-{(\frac{2M}{r})}^{4}dt^{2} + {(\frac{2M}{r})}^{2}dr^{2}
+ M^{2}d{\Omega}^{2}_2\nn
A_0&=&{(\frac{2M}{r})}^{2}
\label{br}
\ena
Since we only consider radial motion, we can put $L^{\apr}_0=0$,
which means we switch off the potential induced from the motion
around $S^2$. When we ignore the contribution from the fermionic
part, the WZ action is reduced to 
\bea
I_{WZ}\sim \int\limits_{{\cal M}_2}A\e_{01}L^0\wedge L^1 \sim
\int\limits_{{\cal M}_2}A{(\frac{2M}{r})}^{2}dt\wedge(\frac{2M}{r})dr
\sim \int dt \frac{A}{r^2}
\label{rp}
\ena
which means the WZ action contains  $r^{-2}$  term that is very
crucial ingredient in the conformal mechanics model ~\cite{AFF}.

In ~\cite{CDK}, the potential is induced by
\bea
V=qA_0 \sim \frac{q}{r^2}
\label{q}
\ena
where q is the electric charge of the superparticle. By 
comparing \p{q} with \p{rp}, we find that the parameter A
can be interpreted as the electric charge of super 0-brane,
and from \p{ma} we can explain the parameter B as the magnetic
charge of super 0-brane. Then Eq.\p{ma}, which relates the mass 
and charges, shows that the action
describes the dynamics of the BPS saturated dyonic superparticle
on Ad$S_2 \times S^2$ background.

Since the non-trivial background fields in Ad$S_2 \times S^2$
vacuum are spacetime metric and RR 2-form flux, the bosonic part of
${{\cal L}_{WZ}}$ (built out
of $L^a$, $L^{\apr}$ and $L^I$) describes, indeed, the bosonic
couplings of 0-brane to the 2-form flux (that is, the required coupling
with $A_0$ field). The action of the super 0-brane contains also the
fermionic terms required to make this coupling supersymmetric
and $\k$-invariant.

From \p{ls}, we know under rescaling of R, ${\cal L}_{DBI}$,
${\cal L}_{WZ1}$ and ${\cal L}_{WZ2}$ transform as
\bea
{\cal L}_{DBI}\rightarrow R^{-1}{\cal L}_{DBI},\;\;
{\cal L}_{WZ1}\rightarrow R^{-1}{\cal L}_{WZ1},\;\;
{\cal L}_{WZ2}\rightarrow R^{-2}{\cal L}_{WZ2}
\ena
which shows ${\cal L}_{WZ2}$ can be ignored in large R, then in
the flat-space limit $R\rightarrow\infty$ the super 0-brane action
is reduced to 
\bea
I^{R\rightarrow\infty}_{0-brane}&=& \int dt\{ -{(A^{2} +
B^{2})}^{1/2}\sqrt{-{({\dot{x}}^{\aha} - 
i{\bar{\t}}^{I}\C^{\aha}{\dot{\t}}^{I})}^2}\nn \;\;
&&  
+ A\e^{IJ}{\bar{\t}}^{I}{\dot{\t}}^{J} + B\e^{IJ}{\bar{\t}}^{I}\C_{5}
{\dot{\t}}^{J}\}
\ena
where $\C^{\aha}$ and $\C_{5}$ are gamma matrices of $SO(1,3)$.

Here we should mention that when we construct WZ action, we could
also include such terms
\bea
{\cal {L}}_{WZ} &=& {\cal H}_{1} + {\cal H}_{2} + \tilde{\cal H}
\ena
with
\bea
\tilde{\cal H}=  \tilde{A}\d^{IJ}L^{\a\alpr
I}(C\c)_{\a\b}C_{\alpr\bepr}\wedge L^{\b\bepr J}+\tilde{B}s^{IJ}L^{\a\alpr
I}C_{\a\b}(C\gapr)_{\alpr\bepr}\wedge L^{\b\bepr J}
\ena
where $s^{IJ}=(1, -1)$. One can easily check
\bea
d\tilde{\cal H}&=&0,\nn 
\d{\tilde{\cal H}}& = & -2d [{\tilde{A}}\d^{IJ}\bar{L}^{I}_{0}
(\c\otimes{1})\d{{\t}^J} + \tilde{B}s^{IJ}\bar{L}^{I}_{0}
({1}\otimes\gapr)\d{\t}^{J}]
\ena
Under rescaling of R, $\tilde{\cal H}\rightarrow R^{-1}
\tilde{\cal H}$, so $\tilde{\cal H}$ can not be ignored in large R 
comparing
with ${\cal {L}}_{DBI}$ and ${\cal {L}}_{WZ1}$. Since $\c$, $\gapr$
are the combinations of gamma matrices of $SO(1,1)$, $SO(2)$
respectively, in $R\rightarrow\infty$ limit
$\tilde{\cal H}$ cannot be expressed in the gamma matrices of 
 $SO(1,3)$, thus the invariance of super 0-brane action under
$SO(1,3)$ cannot be recovered in flat-space limit if  
$\tilde{\cal H}$ was added, so we have to choose $\tilde{A}=\tilde{B}=0$
in the above construction. 

As we know, the Ad$S_2\times S^2$ configuration can arise as the solution in
D=4, N=2 pure supergravity~\cite{GH}, and in extended case coupling
with n vector multiplets~\cite{FKS}. If D=4, N=2 supergravity
model can be obtained from the type IIB superstring compactified on
a Calabi-Yau (CY) threefold (which seem likely in view of the results
of~\cite{KR2}), the above super 0-brane action can be interpreted as
the wrapping of D3-brane action on a supersymmetric cycle ${\cal C}_3$
of the CY manifold ~\cite{BCD}. In type IIB superstring theory,
D3-brane couples to self-dual RR 5-forms, which means D3-brane has
electric and magnetic charges. When D3-brane compactified on
supersymmetric cycle  ${\cal C}_3$, it turns into 0-brane, and
contains selfdual and antiselfdual 2-forms in 4 dimensions~\cite{BCD}. Then
the obtained super 0-brane action should be
interpreted as describing BPS saturated dyonic superparticle
moving on Ad$S^2 \times S^2$, and the electric and magnetic charges
of super 0-brane in 4 dimensions have their origin in 10 dimensions.

\vs{5}
{\Large\bf 4. GS superstring action on $AdS_2 \times S^2$}
\vs{5}

Let us consider GS superstring action on Ad$S_2 \times S^2$,
which is invariant under the local $\k$-symmetry transformation given
by
\bea
\d_{\k}x^a\,\,\, & = & \,\,\, 0, \nn
\d_{\k}x^{\apr}\,\,\, & = & \,\,\,0,\nn
\d_{\k}\t^I &=&2(\c^{a} L_{i}^{a}+ \c\otimes\c^{\apr}L_i^{a'}) \k^{iI}\nn
\d_{\k}(\sqrt{g}g^{ij})&=&-16\sqrt{g}
(P_{-}^{jk}{\bar{L}}_k^{1}\k^{i1}+
P_{+}^{jk}\bar{L}_k^{2}\k^{i2})  
\label{gs}
\ena
with
\bea
P_{\pm}=\frac{1}{2}(g^{ij}\pm \frac{1}{\sqrt{g}}\e^{ij}), \;\;
P_{-}^{ij}\k_{j}^{1} = \k^{i1}, \;\;
P_{+}^{ij}\k_{j}^{2} = \k^{i2}
\label{gs1}
\ena
The GS superstring action on Ad$S_2 \times S^2$, which has global
$SU(1,1\mid 2)$ invariance, should have the form ~\cite{MT1} given by
\bea
I_{GS}&=&-\frac{1}{2}{\int\limits_{\pa {\cal M}_3}d^{2}\sigma\sqrt{-g}
g^{ij}(L^{a}_{i}L^{a}_{j} + L^{\apr}_{i}L^{\apr}_{j})} + 
\int\limits_{{\cal M}_3}{{\cal L}}_{WZ}
\ena
with
\bea
{{\cal L}}_{WZ} = ls^{IJ}(L^{a}\wedge\bar{L}^{I}\c^{a}\wedge L^{J}
+ L^{\apr}\wedge\bar{L}^{I}\c\otimes\c^{\apr}\wedge L^{J})
\ena
One can check by \p{mc}
\bea
d{\cal L}_{WZ} = 0
\ena
so ${{\cal L}}_{WZ}$ is a closed 3-form invariant under $SO(1,1)\times
SO(2)$. The variation of ${{\cal L}}_{WZ}$ takes the following form,
which can be obtained from \p{vari}
\bea
\d{\cal L}_{WZ} = 2ls^{IJ}d(L^{\aha}\wedge\bar{L}^{I}\C^{\aha}\d\t^J)
\ena
By exploiting \p{vari}, \p{gs}, \p{gs1} and requiring $\d I_{GS}=0$, one can
determine $l=1$, and the GS superstring action on Ad$S_{2}\times
S^2$ turns into
\bea
I_{GS}&=&-\frac{1}{2}{\int d^{2}\sigma
\sqrt{g}g^{ij}(L^{a}_{i}L^{a}_{j} +  L^{\apr}_{i}L^{\apr}_{j})}\nn
&&
+ \int\limits_{{\cal M}_3}s^{IJ}[L^{a}\wedge\bar{L}^{I}\c^{a}\wedge
{L}^{J}+L^{\apr}\wedge\bar{L}^{I}\c\otimes\c^{\apr}\wedge{L}^{J}]
\ena
Since the background which we consider is Ad$S_{2}\times
S^2$ with 2-form flux, the constructed WZ term is unique, and
GS superstring action in Ad$S_{2}\times S^2$ with 2-form flux takes
the same form as those on  Ad$S_{5}\times S^5$ and Ad$S_{3}\times S^3$
with RR field.

\vs{5}
{\Large\bf 5. Summary and discussions}
\vs{5}

In the above, we have constructed super
0-brane and GS superstring actions on AdS$_2 \times S^2$ background
with 2-form flux
by supercoset approach. 
For super 0-brane, the action contains
two  parameters A and B, which can be interpreted as
the electric and magnetic charges of the
super 0-brane.
The obtained super 0-brane
action has global $SU(1,1 \mid 2)$ super-invariance, and is
invariant under local  $\k$-symmetry and one-dimensional 
reparametrization invariance. 
The mass of super 0-brane is required to be $m = {(A^{2} + B^{2})}^{1/2}$
by  $\k$-symmetry, which indicates that the action describes
the BPS saturated dyonic superparticle moving on  AdS$_2 \times S^2$ 
background with 2-form flux. Also the bosonic part of the
WZ action  contains the bosonic couplings of 0-brane to the 2-form
flux background, which is crucial ingredient in the
conformal mechanics model~\cite{AFF}. For GS superstring, we
find that the string  action  on AdS$_2 \times S^2$ with 2-form flux
takes the same form as those in AdS$_3 \times S^3$ and
AdS$_5 \times S^5$ with RR field background.

By the supercoset approach, we can find the closed form expression
for the supervielbein by solving the Maurer-Cartan equation (for
AdS$_5 \times S^5$, such an expression has been found in~\cite{KRR}).
In~\cite{PST}, it was shown that the supersolvable algebra gauge
and the Killing spinor gauge for fixing the $\k$-symmetry of brane
actions in $AdS_{p+2}\times S^{D-p-2}$ backgrounds are incompatible
with supersymmetric static solutions of worldvolume brane equations
of motion, then it is interesting to see how to do gauge-fixing of 
$\k$-symmetry properly for super 0-brane action on 
AdS$_2 \times S^2$, which would shed light on the similar problems
on other complicated backgrounds.

If we consider a restriction on the full dynamics in which the
particle is assumed to move within an equatorial plane (this
restriction corresponds to a reduction of the superconformal
symmetry to the subgroup of $SU(1,1\mid 2)$:  $SU(1,1\mid 2)\supset 
SU(1,1\mid 1)\cong OSp(2\mid 2)$), it would be interesting to study
how to reduce our model to that in ~\cite{AP}, from which we could
draw some lessons on the quantization of super 0-brane on 
AdS$_2 \times S^2$.

It was argued that the large n-particle 
$SU(1,1\mid 2)$ superconformal Calogero model 
would provide a microscopic description of the extremal
RN black hole (at least near the horizon)~\cite{GT}. Then we
expect that the n-particle
$SU(1,1\mid 2)$ superconformal Calogero model will probably give us
some hints on the non-abelian generalization of the abelian
0-brane action that we have found. We hope to return those
issues in near future.

\section*{Acknowledgement} 

I would like to thank E. Bergshoeff, R.R. Metsaev, 
N. Ohta and Jaemo Park for valuable discussions, and
D. Sorokin for helpful comment.

\newcommand{\NP}[1]{Nucl.\ Phys.\ {\bf #1}}
\newcommand{\AP}[1]{Ann.\ Phys.\ {\bf #1}}
\newcommand{\PL}[1]{Phys.\ Lett.\ {\bf #1}}
\newcommand{\CQG}[1]{Class. Quant. Gravity {\bf #1}}
\newcommand{\CMP}[1]{Comm.\ Math.\ Phys.\ {\bf #1}}
\newcommand{\PR}[1]{Phys.\ Rev.\ {\bf #1}}
\newcommand{\PRL}[1]{Phys.\ Rev.\ Lett.\ {\bf #1}}
\newcommand{\PRE}[1]{Phys.\ Rep.\ {\bf #1}}
\newcommand{\PTP}[1]{Prog.\ Theor.\ Phys.\ {\bf #1}}
\newcommand{\PTPS}[1]{Prog.\ Theor.\ Phys.\ Suppl.\ {\bf #1}}
\newcommand{\MPL}[1]{Mod.\ Phys.\ Lett.\ {\bf #1}}
\newcommand{\IJMP}[1]{Int.\ Jour.\ Mod.\ Phys.\ {\bf #1}}
\newcommand{\JHEP}[1]{J.\ High\ Energy\ Phys.\ {\bf #1}}
\newcommand{\JP}[1]{Jour.\ Phys.\ {\bf #1}}

\end{document}